# Numerical Solution of the 1D-Schrödinger Equation with Pseudo-Delta Barrier Using Numerov Method

[1]S. D. G. Martinz and [2]R. V. Ramos
samueldgm@fisica.ufc.br   rubens.viana@pq.cnpq.br

[1]*Federal Institute of Education, Science and Technology of Ceara, Fortaleza-Ce, Brazil.*
[2]*Lab. of Quantum Information Technology, Department of Teleinformatic Engineering – Federal University of Ceara - DETI/UFC, C.P. 6007 – Campus do Pici - 60455-970 Fortaleza-Ce, Brazil.*

In this work, aiming to solve numerically the Schrödinger equation with a Dirac delta function potential, we use the Numerov method to solve the time independent 1D-Schrödinger equation with potentials of the form $V(x) + \alpha\delta_p(x)$, where $\delta_p(x)$ is a pseudo-delta function, a very high and thin barrier. The numerical results show good agreement with analytical results found in the literature. Furthermore, we show the numerical solutions of a system formed by three delta function potentials inside of an infinite quantum well and the harmonic potential with position dependent mass and a delta barrier in the center.

## 1. Introduction

The solutions of the Schrödinger equation for different potentials have attracted much interest since the early days of quantum mechanics. Nowadays that interest has been renewed due to the design and fabrication of nanodevices. An interesting situation occurs when the potential includes a delta function or its derivative. Some analytical solutions have been provided for special cases [1-4] and a good review can be found in [5]. In order to obtain more results with arbitrary potential including the delta function as barrier, a numerical method could be used. In this direction, the present work uses the Numerov method to solve numerically the time independent 1-D Schrödinger equation with potentials of the form $V(x) + \alpha\delta_p(x)$, where $\delta_p(x)$ is a pseudo-delta function, a very high and thin barrier. Such potential is used here as an approximation of potentials with Dirac delta function. Our results show a good agreement with known analytical solutions, as well we consider new potentials not found in the literature: three delta barriers inside of an infinite quantum well and the delta function in the center of the harmonic potential in a 1D position mass dependent Schrödinger equation.

## 2. The Numerov method and the numerical solution of the Schrödinger equation with pseudo-delta potential

In order to find the solutions of the time independent 1D-Schrödinger equation, one has to solve

$$\frac{d^2\psi(x)}{dx^2} = -\frac{2m}{\hbar^2}\left[E - V(x)\right]\psi(x) \qquad (1)$$

together with the appropriated boundary conditions. Its discretization using the Numerov method is [6]

$$\left[-\frac{\hbar^2}{2m}B^{-1}A+V\right]\psi = E\psi \tag{2}$$

$$\psi = \begin{bmatrix} \ldots & \psi(x_{k-1}) & \psi(x_k) & \psi(x_{k+1}) & \ldots \end{bmatrix}^T \tag{3}$$

$$A = (I_{-1} - 2I_0 + I_1)/(\Delta x)^2 \tag{4}$$

$$B = (I_{-1} + 10I_0 + I_1)/12 \tag{5}$$

$$V = \begin{bmatrix} V(x_1) & 0 & 0 \\ 0 & \ddots & 0 \\ 0 & 0 & V(x_n) \end{bmatrix}. \tag{6}$$

In (2)-(6) $I_k$, $k = 0$, -1, and 1, is a matrix of 1s along the $k$th diagonal and zeros elsewhere. Here, we are interested in the solutions of

$$-\frac{\hbar^2}{2m}\frac{d^2\psi}{dx^2} + \left[V(x) + \alpha\delta_p(x)\right]\psi = E\psi \tag{7}$$

where $\delta_p(x)$ is a very thin and high barrier located around the point $x = 0$.

Before showing the numerical results of (7) using (2)-(6), we make a brief review of some analytical results about the eigenfunctions and energies when the potential includes a delta function. The first result, easily found in textbooks of quantum mechanics, is the discontinuity of the first derivative of the eigenfunctions in the position of the delta function. For a delta function in $x = 0$ and strength parameter $\alpha$, one has

$$\left[\frac{d\psi}{dx}\bigg|_{x\to 0^-} - \frac{d\psi}{dx}\bigg|_{x\to 0^+}\right] = \frac{2m\alpha}{\hbar^2}\psi(0). \tag{8}$$

From (8) one can note that if $\psi(0) = 0$, then the particle does not "see" the delta function and the first derivative is continuous. Now, let us consider the eigenfunction $\phi_n(x)$ with associated energy $E_n$ a solution of the Schrödinger equation with potential $V(x)$

$$-\frac{\hbar^2}{2m}\frac{d^2\phi_n}{dx^2} + V(x)\phi_n = E_n\phi_n. \tag{9}$$

The solution of the Schrödinger equation with potential $V(x) + \alpha\delta(x)$, $\psi(x)$, can be written in the $\phi_n(x)$ basis as $\psi(x) = \sum_n c_n \phi_n(x)$. Thus, substituting in the Schrödinger equation one gets

$$-\frac{\hbar^2}{2m}\frac{d^2\sum_n c_n\phi_n}{dx^2}+\left[V(x)+\alpha\delta(x)\right]\sum_n c_n\phi_n = E\sum_n c_n\phi_n \tag{10}$$

$$\Rightarrow \sum_n c_n\frac{-\hbar^2}{2m}\frac{d^2\phi_n}{dx^2}+V(x)\phi_n+\sum_n c_n\left[\alpha\delta(x)-E\right]\phi_n = 0 \tag{11}$$

$$\Rightarrow \sum_n c_n\left[\alpha\delta(x)+(E_n-E)\right]\phi_n = 0. \tag{12}$$

Now, taking the inner product of (12) with $\phi_m^*(x)$, one has

$$\int_{-\infty}^{\infty}\sum_n c_n\left[\alpha\delta(x)+(E_n-E)\right]\phi_n\phi_m^*dx = 0 \tag{13}$$

$$\Rightarrow \sum_n c_n\alpha\int_{-\infty}^{\infty}\delta(x)\phi_n\phi_m^*dx+\sum_n c_n(E_n-E)\int_{-\infty}^{\infty}\phi_n\phi_m^*dx = 0 \tag{14}$$

$$\Rightarrow \sum_n c_n\alpha\phi_n(0)\phi_m^*(0)+c_m(E_m-E) = 0 \tag{15}$$

$$\Rightarrow \alpha\phi_m^*(0)\psi(0) = c_m(E-E_m) \Rightarrow c_m = \frac{\alpha\phi_m^*(0)\psi(0)}{(E-E_m)}. \tag{16}$$

Hence,

$$\psi(x) = \sum_n \frac{\alpha\phi_n^*(0)\psi(0)}{(E-E_n)}\phi_n(x) \tag{17}$$

Finally, using (17) for calculating $\psi(0)$, one gets

$$\frac{1}{\alpha} = \sum_n \frac{|\phi_n(0)|^2}{(E-E_n)} \tag{18}$$

Clearly (18) imposes a restriction on the allowed energy values. In the first order approximation one has $E \sim E_n + \alpha|\phi_n(0)|^2$ [5]. Now, returning to (17), the normalization of the quantum state $\psi(x)$ requires that

$$\int_{-\infty}^{\infty}|\psi(x)|^2 dx = 1 = \int_{-\infty}^{\infty}\sum_n \frac{\alpha\phi_n^*(0)\psi(0)}{(E-E_n)}\phi_n(x)\sum_m \frac{\alpha\phi_m(0)\psi^*(0)}{(E-E_m)}\phi_m^*(x)dx \tag{19}$$

$$= \alpha^2|\psi(0)|^2\sum_{n,m}\frac{\phi_n^*(0)\phi_m(0)}{(E-E_n)(E-E_m)}\int_{-\infty}^{\infty}\phi_n(x)\phi_m^*(x)dx \tag{20}$$

$$\Rightarrow \alpha^2|\psi(0)|^2\sum_n \frac{|\phi_n^*(0)|^2}{(E-E_n)^2} = 1 \tag{21}$$

For some famous quantum systems like infinite and finite quantum wells and the harmonic potential, one has $\phi_n(0) = 0$ for half of the possible values of $n$. Hence, from (21), one has that if $E = E_n$ then $\psi(x) = \phi_n(x)$ with $\phi_n(0) = 0$ and vice-versa.

The first issue that one faces when tries to solve numerically the Schrödinger equation with a delta barrier is the discretization of the delta function. The obvious answer is to use some approximation of the delta function. For example, one can use one of the following three well known approximations of the Dirac delta function as a pseudo-delta function

$$\delta_p^R(x) = \begin{cases} 1/(2\varepsilon) & -\varepsilon \leq x \leq \varepsilon \\ 0 & x < -\varepsilon \text{ or } x > \varepsilon \end{cases} \quad (22)$$

$$\delta_p^G(x) = \frac{1}{2\sqrt{\pi\varepsilon}} \exp\left(-\frac{x^2}{4\varepsilon}\right) \quad (23)$$

$$\delta_p^P(x) = \frac{1}{\pi} \frac{\varepsilon}{x^2 + \varepsilon^2}. \quad (24)$$

The Dirac delta function is obtained from (22), (23) or (24) when $\varepsilon \to 0$. In order to choose one of them for our simulations we consider the following criterion: Equation (2) can be rewritten as $[S + \delta_p]\psi = E\psi$, where $S$ is the Hamiltonian without delta function, $S = (-\hbar^2/2m)B^{-1}A + V$. Its solution requires that $\det(S + \delta_p - EI) = 0$, where $E$ is an eigenvalue of $(S + \delta_p)$, $\delta_p$ is a diagonal matrix whose entries are the discrete values of the pseudo-delta function chosen and $I$ is the identity matrix. Now, the following modification can be implemented without changing the result of the determinant: $\det(S - E_nI + \delta_p - (E-E_n)I) = 0$, where $E_n$ is an eigenvalue of $S$. Using the Minkowski determinant theorem, one has $[\det(S - E_nI + \delta_p - (E-E_n)I)]^{1/k} \geq [\det(S - E_nI)]^{1/k} + [\det(\delta_p - (E-E_n)I)]^{1/k}$, which holds true if $(S - E_nI)$ and $(\delta_p - (E-E_n)I)$ are non-negative $k \times k$ Hermitean matrices. This is the case when $E = E_n$ since the eigenvalues of the matrix $\delta_p$ are exactly the elements of $\delta_p$ that, according to (22)-(24), are always non-negatives. Thus, using the Minkowski determinant theorem one has $0 \geq 0 + [\det(\delta_p)]^{1/k}$. The equality is true only if at least one element of $\delta_p$ is zero. Although this is a very good approximation for (23) and (24) when $\varepsilon$ is large, it is really true for (22) even when a not so large value of $\varepsilon$ is used. Hence, our simulations will use only (22) as a pseudo-delta function.

At last, since the pseudo-delta barrier has finite height, the strength parameter $\alpha$ is absorbed in the barrier height.

## 3. The numerical solution of the mass independent position Schrodinger equation with delta potential

We start by considering the delta function in $x = 0$ in the middle of an infinite quantum well ($-10$nm $< x < 10$nm) with width equal to $2L = 20$nm. There are two types of solutions for this case [7,8]. The odd solutions are exactly the same odd solutions of the quantum infinite well without the delta function and the energies are also the same, as foreseen by (21). On the other hand, the allowed energies for the even solutions are those that satisfy the condition [8]

$$\tan\left(\frac{\sqrt{2mE}}{\hbar}L\right) = -\frac{\hbar^2}{m\alpha}\left(\frac{\sqrt{2mE}}{\hbar}\right). \tag{25}$$

Hence, the quality of the numerical results is checked by testing whether found solutions of the even modes satisfy equations (8) and (25). For this system, Fig. 1 shows the squared modulus of the eigenfunctions $\psi_1$ and $\psi_3$ and their first derivatives with the expected discontinuities at $x = 0$. In Fig. 2 it is shown the energies for the first thirty modes while in Fig. 3 it is shown how good the energy values found numerically satisfy (25), for the first twenty energies' values of even modes. At last, in Fig. 4 it is shown how good the boundary condition given by (8) is satisfied by the derivatives of the first 15 even modes. The particle's mass is $0.067m_e$. In all figures the eigenfunctions and their first derivatives are multiplied and displaced by a constant factor in order to permit their visualization inside of the potential function profile.

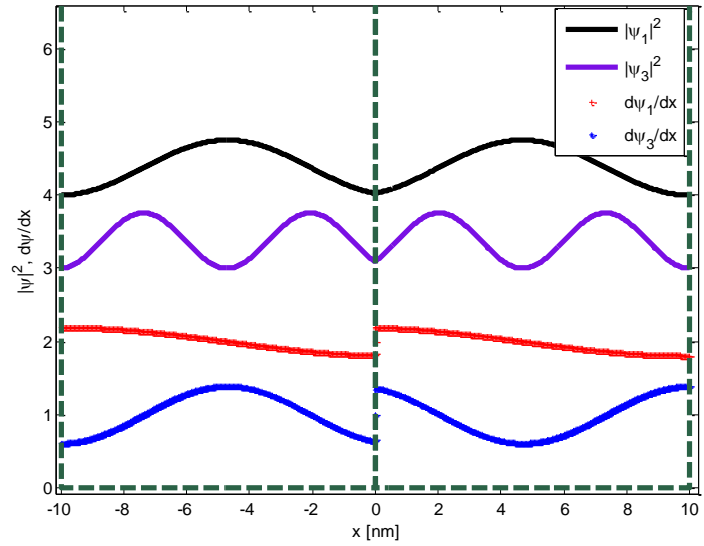

Fig. 1 – The squared modulus and the first derivatives of the eigenfunctions $\psi_1$ and $\psi_3$.

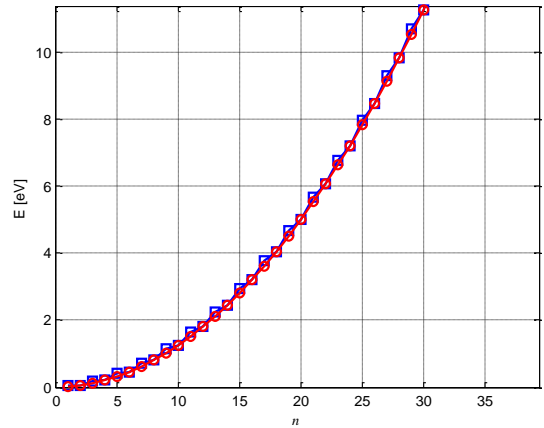

Fig. 2 – Energies for the first thirty modes. 'ball'- Infinite quantum well without delta function. 'square' - Infinite quantum well with delta function in the middle.

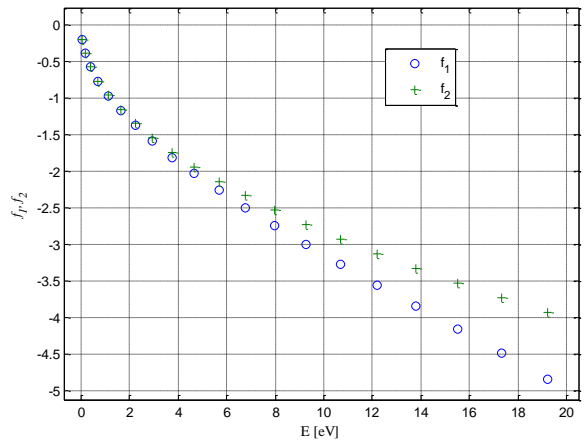

Fig. 3 - Comparison between $f_1 = \tan(kL)$ and $f_2 = -\hbar^2 k/(m\alpha)$ (Eq. (25)) using the energy values found numerically for the first twenty even modes. $k = (2mE)^{1/2}/\hbar$ (infinite quantum well with delta barrier in the middle).

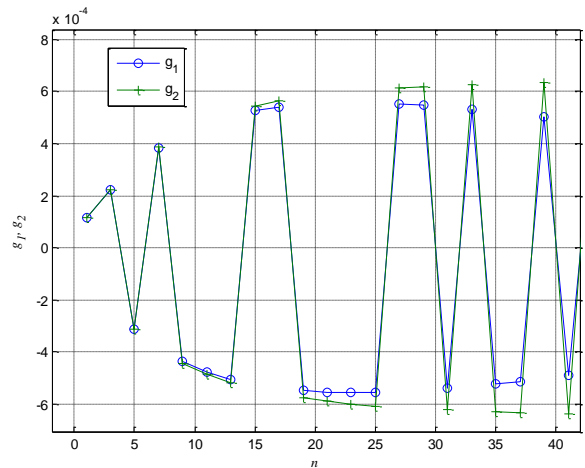

Fig. 4 – Comparison between $g_1 = d\psi/dx|_{0-} - d\psi/dx|_{0+}$ and $g_2 = (2m\alpha/\hbar^2)\psi(0)$ for the first 20 even modes (infinite quantum well with delta barrier in the middle).

Similarly, for the finite quantum well with a delta barrier in the middle, one can see in Fig. 5 the squared modulus of the eigenfunctions $\psi_1$ and $\psi_3$ while their derivatives with the expected discontinuities at $x = 0$ are shown in Fig. 6. In Fig. 7 it is shown the energies for the first nineteen modes while in Fig. 8 it is shown how good the boundary condition given by (8) is satisfied by the derivatives of the first sixteen even modes.

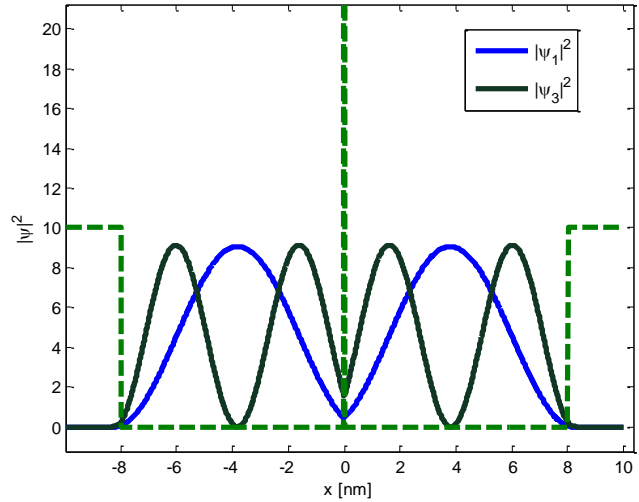

Fig. 5 – The squared modulus of the eigenfunctions $\psi_1$ and $\psi_3$.

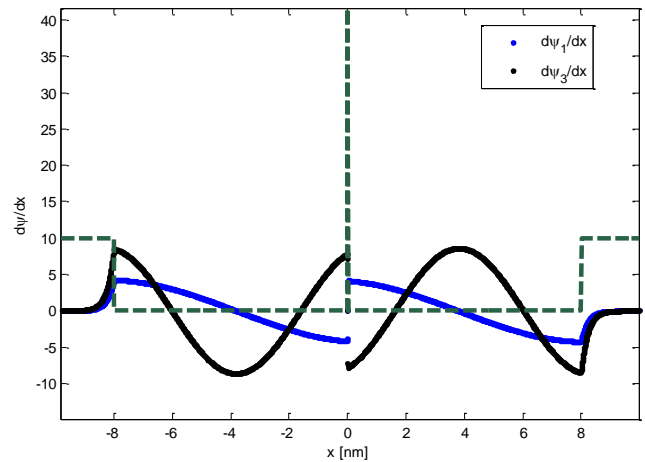

Fig. 6 – Derivatives of $\psi_1$ and $\psi_3$ versus $x$.

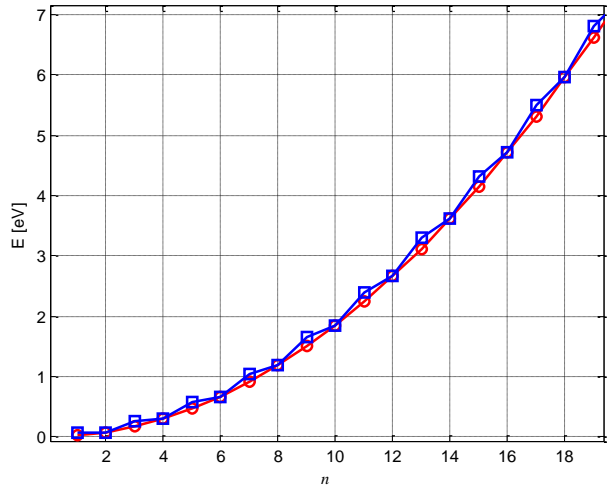

Fig. 7 – Energies for the first nineteen modes. 'ball'- finite quantum well. 'square' - finite quantum well with delta function in the middle.

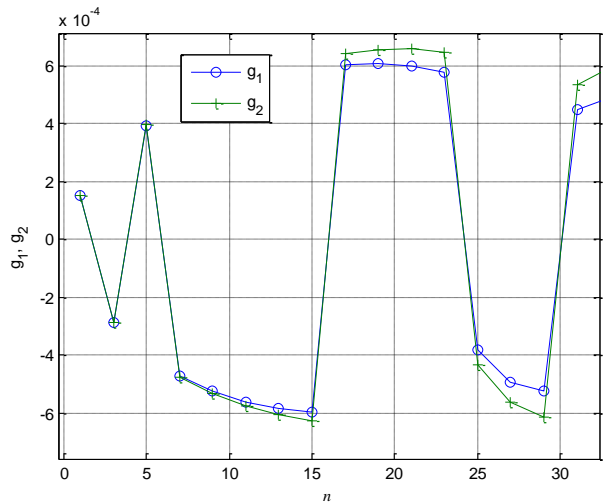

Fig. 8 – Comparison between $g_1 = d\psi/dx|_{0-} - d\psi/dx|_{0+}$ and $g_2 = (2m\alpha/\hbar^2)\psi(0)$ for the first sixteen even modes (finite quantum well with delta barrier in the middle).

Now, the harmonic potential (the frequency of oscillation is $10^{15}$Hz) with delta function in the middle is considered. In Fig. 9 one can see the squared modulus of the eigenfunctions $\psi_1$ for both cases with (II) and without (I) delta function while in Fig. 10 it is shown the same for $\psi_3$. The derivatives of $\psi_1$ and $\psi_3$ are shown in Fig. 11, the energies for the first 24 modes, with and without delta function, are given in Fig. 12 and, at last, Fig. 13 shows how good the boundary condition given by (8) is satisfied by the derivatives of the first sixteen even modes.

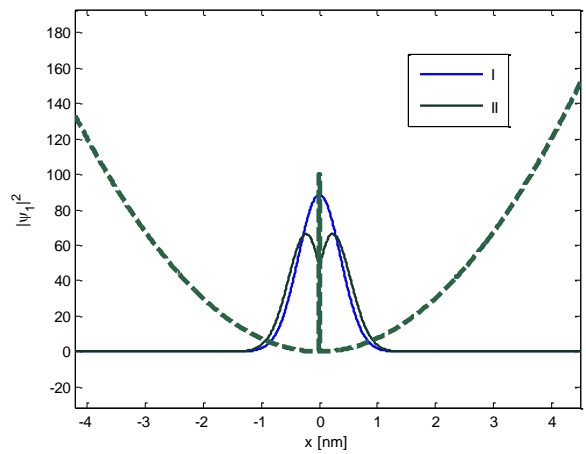

Fig. 9 - The squared modulus of the eigenfunction $\psi_1$ (I – without delta function; II – with delta function at $x = 0$) for the harmonic potential with frequency $10^{15}$Hz.

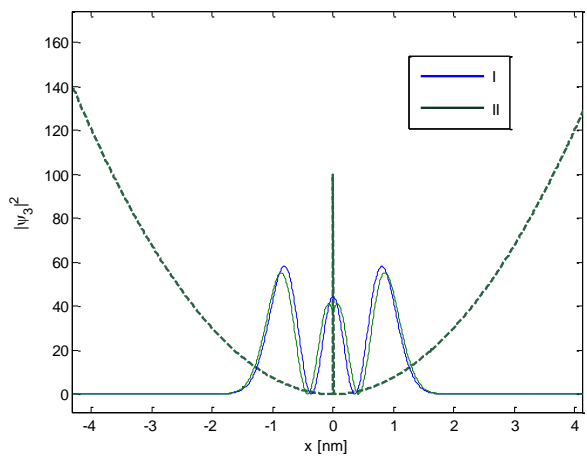

Fig. 10 - The squared modulus of the eigenfunction $\psi_3$ (I – without delta function; II – with delta function at $x = 0$) for the harmonic potential with frequency of oscillation $10^{15}$Hz.

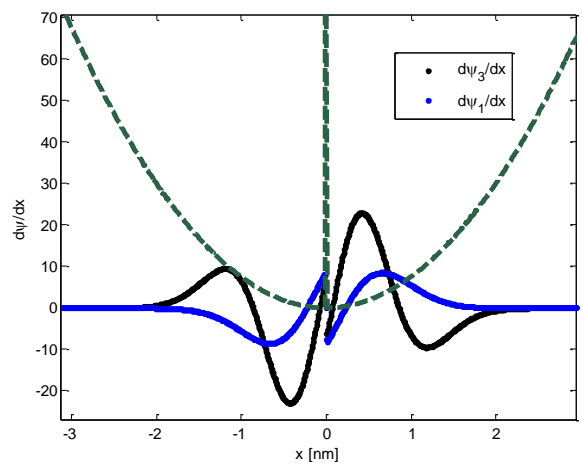

Fig. 11 – Derivatives of $\psi_1$ and $\psi_3$ versus $x$.

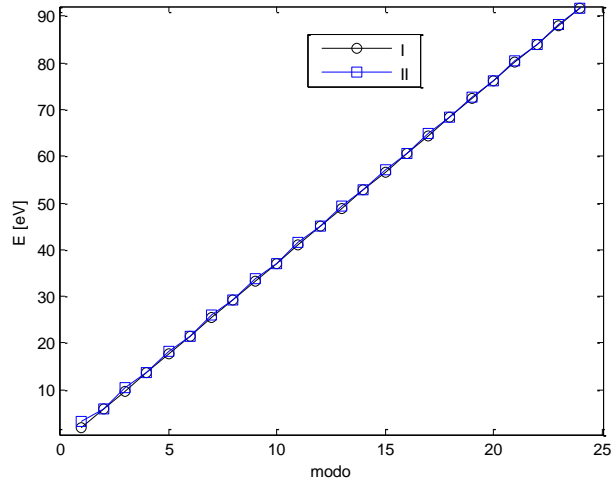

Fig. 12 – Energies for the first 24 modes. I- harmonic potential without delta function. II - harmonic potential with delta function in the middle.

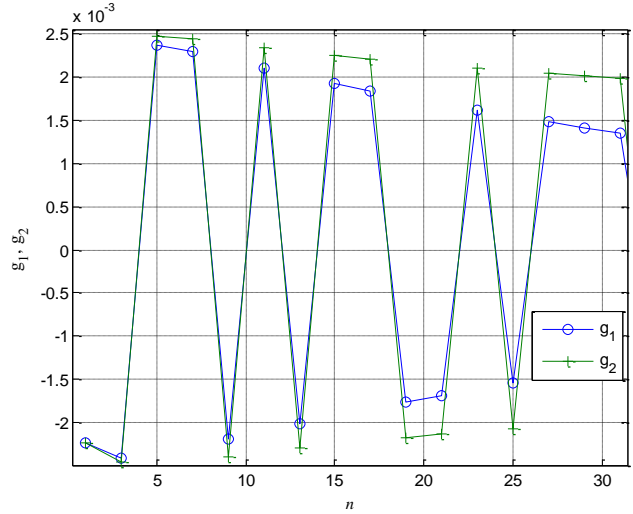

Fig. 13 – Comparison between $g_1 = d\psi/dx|_{0^+} - d\psi/dx|_{0^-}$ and $g_2 = (2m\alpha/\hbar^2)\psi(0)$ for the first sixteen even modes (harmonic potential with delta barrier in the middle).

At last, we consider an infinite quantum well of width equal to 21nm ($0 \leq x \leq$ 21nm) with three delta barriers, located at $x = 4.5$nm, $x = 10.5$nm and $x = 16.5$nm. The squared modulus and the first derivatives of the eigenfunctions $\psi_1$ and $\psi_2$ can be seen in Fig. 14. Similar plots for $\psi_9$ and $\psi_{10}$ are shown in Fig. 15. The allowed energies' values for the first 25 modes of four different situations are shown in Fig. 16: I) Three delta barriers. II) Only the two lateral delta barriers. III) Only the central delta barrier. IV) Infinite well without any delta barrier.

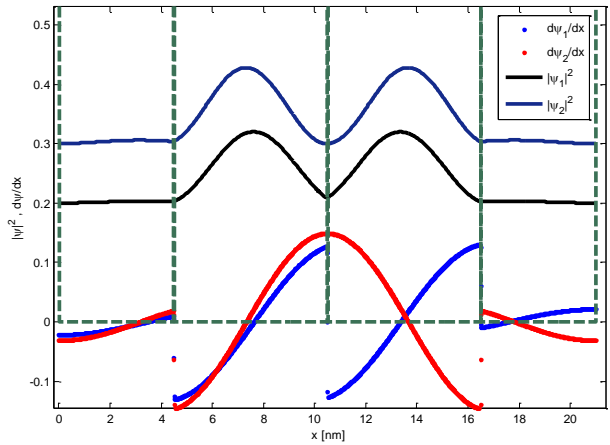

Fig. 14 - Squared modulus and first derivatives of the eigenfunctions $\psi_1$ and $\psi_2$.

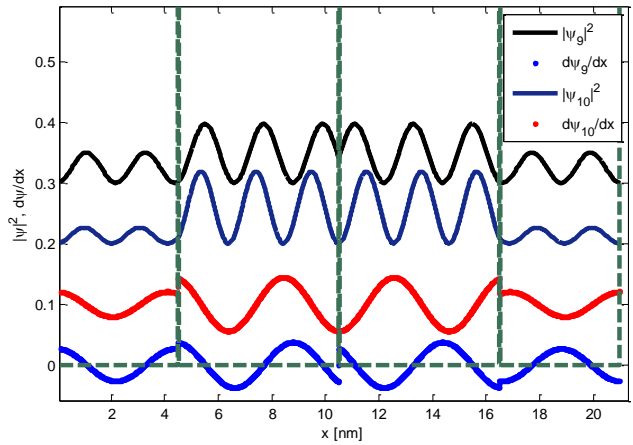

Fig. 15 - Squared modulus and first derivatives of the eigenfunctions $\psi_9$ and $\psi_{10}$.

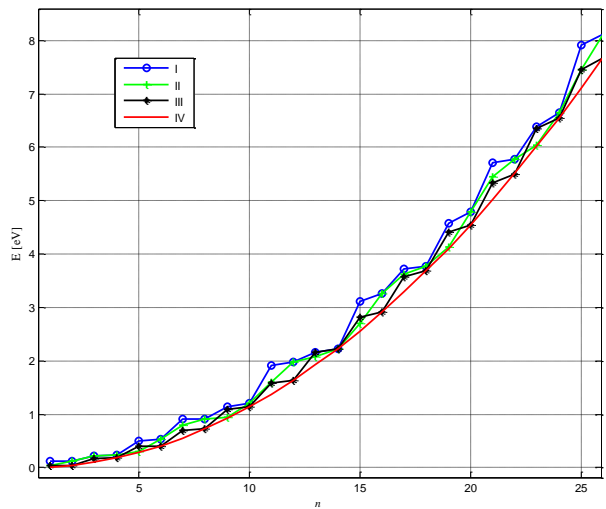

Fig. 16 – Energies for the first 25 modes (infinite quantum well with three delta barriers).

The odd modes do not have discontinuity of the first derivative at the central barrier localization, but there are discontinuities at the lateral barriers positions. The even modes have discontinuities of the first derivative in all three delta barriers. One can also note in Fig. 16 that, as expected, the parabolic behavior of the energy is broken by the delta barriers. Moreover, the larger the number of delta barriers, the larger are the energies' values.

## 4. The numerical solution of the mass dependent position Schrodinger equation with delta potential

In the cases where the particle's mass depends on the position, mass and momentum operators do not commute. Thus, one has to change the kinetic energy operator $T$ in order to get a Hermitean operator $H = T + V$ for the Hamiltonian. The operator proposed by Von Ross [9] was

$$T(x) = \frac{1}{4}\left(m^\alpha p m^\beta p m^\gamma + m^\gamma p m^\beta p m^\alpha\right) \quad (26)$$

$$\alpha + \beta + \gamma = -1. \quad (27)$$

Substituting the operator $p=-i\hbar d/dx$ in (26), the Schrödinger equation can be rewritten as [10]

$$\frac{d^2\psi}{dx^2} - \frac{m'}{m}\frac{d\psi}{dx} + \left[\frac{1}{2}\left(r\frac{m''}{m} - s\frac{m'^2}{m^2}\right) + \frac{2m}{\hbar^2}(E - V(x))\right]\psi = 0. \quad (28)$$

In (28), $r = \alpha + \gamma$, $s = \alpha(\gamma + 2) - \gamma(\alpha + 2)$. Moreover, $m' = dm/dx$ and $m'' = d^2m/dx^2$. Now, making the substitution $\psi(x) = \sqrt{m(x)}\xi(x)$ in (28), the Schrödinger equation can be simplified to

$$\frac{\hbar^2}{2m_e}\frac{d^2\xi}{dx^2} = m_{rel}(x)\left[E - V_{eff}(x)\right]\xi = 0 \quad (29)$$

$$V_{eff}(x) = V(x) - \left[\hbar^2(1+r)\frac{m''}{4m^2} - \hbar^2\left(\frac{3}{4} + \frac{s}{2}\right)\frac{m'^2}{2m^3}\right]. \quad (30)$$

In (29) $m_{rel}(x) = m(x)/m_e$. Its discretization using the Numerov method is

$$\left[-\frac{\hbar^2}{2m_e}M^{-1}A+V\right]\psi = E\psi \tag{31}$$

$$M = (M_{-1}+10M_0+M_1)/12 \tag{32}$$

$$M_{-1} = \begin{bmatrix} 0 & & & \\ m_{rel}(x_1) & 0 & & \\ & \ddots & \ddots & \\ & & m_{rel}(x_{n-1}) & 0 \end{bmatrix}, \quad M_1 = \begin{bmatrix} 0 & m_{rel}(x_2) & & \\ & 0 & \ddots & \\ & & \ddots & m_{rel}(x_n) \\ & & & 0 \end{bmatrix} \tag{33}$$

In (31) and (32) the matrix $A$ is given by (4), $V$ is a diagonal matrix whose entries are the values of $V_{eff}(x)$ and $M_0$ is a diagonal matrix whose entries are the values of $m_{rel}(x)$. Here we are going to use $r = -1$ and $s = -3/2$ ($\alpha = \gamma = -1/2, \beta = 0$), what implies in $V_{eff}(x) = V(x)$, and the mass position dependent Schrödinger equation is simply given by

$$\frac{\hbar^2}{2m_e}\frac{d^2\xi}{dx^2} = m_{rel}(x)\left[E-V(x)\right]\xi = 0. \tag{34}$$

Here, we are going to solve (34) for the harmonic potential with a delta barrier in the center. Initially we consider the following expression for the mass,

$$m_{rel}(x) = 0.0665 + 0.0835x^2. \tag{35}$$

The squared modulus and first derivatives of the eigenfucntions $\psi_1$, $\psi_3$ and $\psi_5$ are shown in Figs. 17 and 18, respectively, while a comparison of the allowed energies for the first eleven modes for harmonic potentials with and without delta barrier and variable mass is shown in Fig. 19.

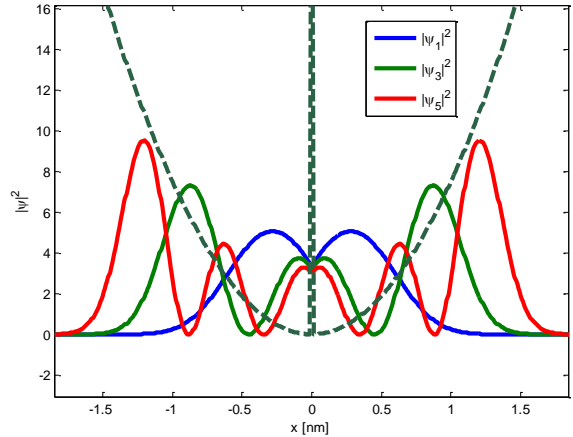

Fig. 17 - Squared modulus of the eigenfunctions $\psi_1, \psi_3$ and $\psi_5$.

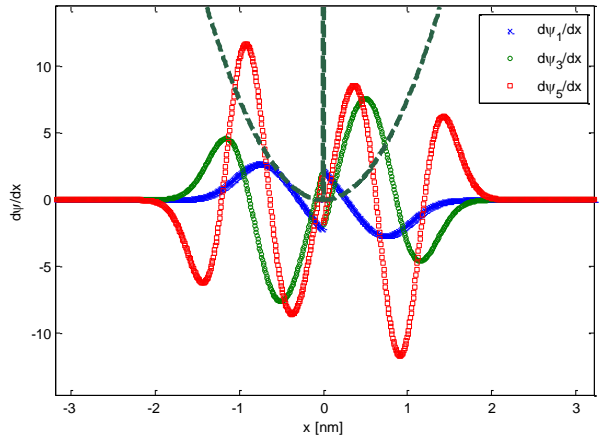

Fig. 18 - First derivatives of the eigenfunctions $\psi_1, \psi_3$ and $\psi_5$.

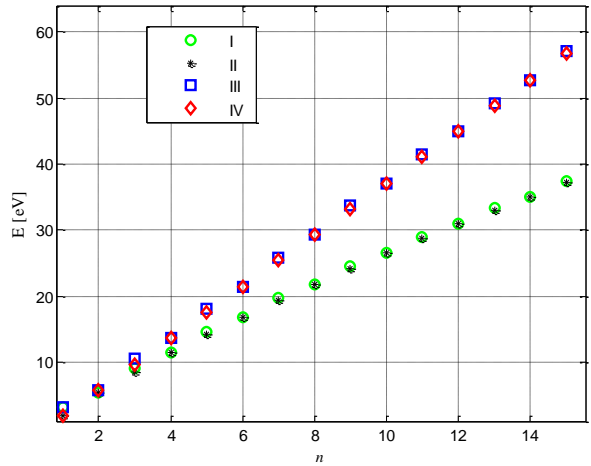

Fig. 19 – Energies for the first eleven modes of the harmonic potential with and without delta barrier and mass variation. I – Delta function and variable mass; II – Variable mass; III) Delta function and constant mass; IV) Constant mass.

At last, we consider a Gaussian profile for the mass, as given by (36),

$$m_{rel}(x) = 1 + 0.67 \exp(-x^2). \qquad (36)$$

The squared modulus and first derivatives of the eigenfucntions $\psi_1$, and $\psi_3$ are shown in Fig. 20, while a comparison of the allowed energies for the first ten modes for harmonic potentials with and without delta function and mass variation is shown in Fig. 21.

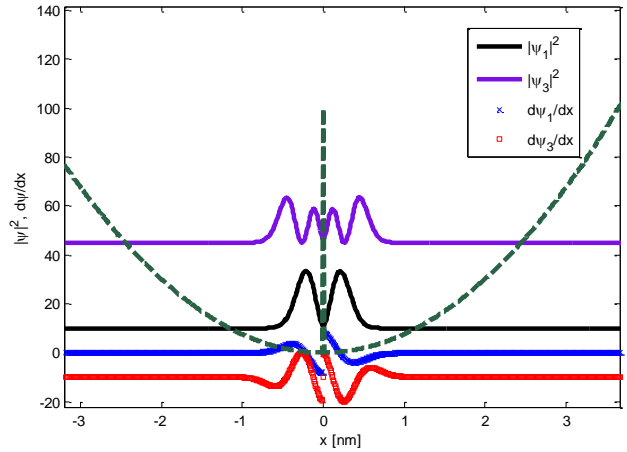

Fig. 20 - Squared modulus and first derivatives of the eigenfunctions $\psi_1$ and $\psi_3$.

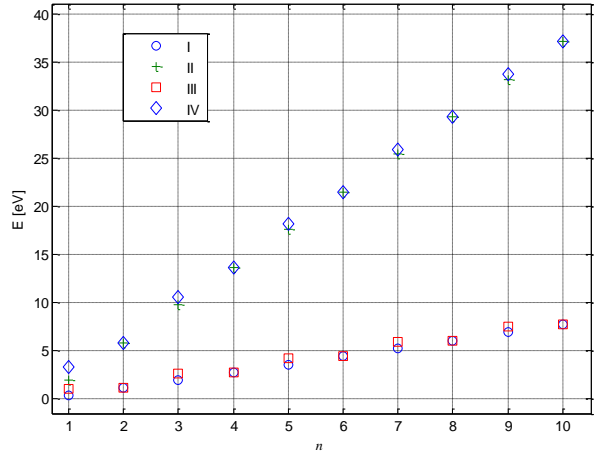

Fig. 21 – Energies for the first ten modes of the harmonic potential with and without delta barrier and mass variation. I – Delta function and constant mass; II – Constant mass; III) Delta barrier and variable mass; IV) Delta barrier and variable mass.

Clearly, the energy of the modes grows more slowly when the mass depends on the position. Furthermore, the energy does not change linearly with the mode number as happen in the constant mass without delta barrier case.

## 5. Conclusions

The numerical solution, via Numerov method, of the Schrödinger equation with delta barrier by simulating the last by a very thin and narrow rectangular barrier can provide good results. Due to the finite height and width of the pseudo-delta barrier, the results for the first two dozen modes are more reliable. In general, the accuracy can be checked seeing how good the boundary condition (8) is satisfied by the numerical solutions found, as it was done in Figs 4, 8 and 13. In all the cases considered the presence of the delta barrier changes the even modes and their energies. The odd modes are changed only in the quantum well with three delta barriers, since in this case the two lateral barriers are located in positions where the wavefunction is not zero. These changes are larger for low order modes (since the pseudo-delta barrier has finite height). In particular, the smooth behavior of the allowed energies (parabolic for the infinite quantum well and linear for the harmonic potential) is broken by the presence of the delta barriers.

## Acknowledgments

This work was supported by the Brazilian agency CNPq via Grant no. 303514/2008-6. Also, this work was performed as part of the Brazilian National Institute of Science and Technology for Quantum Information.